\documentclass[prd,showpacs,showkeys,floatfix,nofootinbib,fleqn,tightenlines,
preprint]{revtex4}

\usepackage{amsmath}\usepackage{amsfonts}
\usepackage{latexsym}\usepackage{graphicx}
\usepackage{mathrsfs}\usepackage{dcolumn}

\newcommand{\beq}{\begin{equation}}
\newcommand{\eeq}{\end{equation}}
\newcommand{\beqa}{\begin{eqnarray}}
\newcommand{\eeqa}{\end{eqnarray}}
\newcommand{\bsubeqs}{\begin{subequations}}
\newcommand{\esubeqs}{\end{subequations}}

\begin{document}
\noindent Phys. Rev. D 85, 016011 (2012)
\hspace*{\fill}  arXiv:1110.2146\vspace*{2\baselineskip}%
%
%
%
\title{Superluminal neutrino, flavor, and relativity}

\author{F.R. Klinkhamer}
\email{frans.klinkhamer@kit.edu} \affiliation{ \mbox{Institute for
Theoretical Physics, University of Karlsruhe,}
\mbox{Karlsruhe Institute of Technology, 76128 Karlsruhe, Germany}\\}

\begin{abstract}
\vspace*{.125\baselineskip}\noindent  
Modified neutrino dispersion relations, which still obey the
relativity principle, can have both a superluminal
(muon-type) neutrino and a luminal (electron-type) neutrino,
as long as neutrino-mass effects can be neglected.
The idea is to allow for flavor-dependent deformed Lorentz
transformations and an appropriate hierarchy of energy scales.
If OPERA's result on a superluminal velocity of the muon-neutrino
is confirmed, the model has a matching superluminal velocity
of the corresponding charged lepton, the muon, at equal particle
energy. Assuming that this model is not already ruled out,
new TeV-scale effects in the muon sector are predicted.
Also discussed is a different model with a superluminal
sterile neutrino propagating in the usual 4 spacetime dimensions.
\vspace*{2\baselineskip}
\end{abstract}


\maketitle

\section{Introduction}
\label{sec:Introduction}

Let us be radical and take OPERA's result~\cite{OPERA2011}
at face value (combining the quoted statistical and systematic errors
in quadrature):
\beq\label{eq:OPERA-result}
\Big(\big[v_{\,\nu_\mu}\,
\big]_{\langle c\,|\mathbf{p}| \rangle\, = \,17\;\text{GeV}}^\text{(exp)}
\,- c\,\Big)\Big/c
=
(2.4\pm 0.4) \times 10^{-5}\,,
\eeq
with $c$ the velocity of light \textit{in vacuo}
and $v$ the inferred time-of-flight velocity of the neutrino.
A similar value, but with larger errors, has been reported
earlier by MINOS~\cite{MINOS2007}.
We are, of course, well aware of the reservations
expressed regarding the claimed OPERA result
(some papers are listed in Ref.~\cite{OPERA2011-doubts}),
but the purpose, here, is to set a problem which is
as difficult as possible, taking the surprisingly
large magnitude indicated by  \eqref{eq:OPERA-result} as given.

Using the observed neutrino burst from supernova SN1987a~\cite{SN1987a},
interpreted as a burst of electron-type antineutrinos,
we also have the following bound~\cite{Longo1987}:
\beqa\label{eq:SN-bound}
\Big|\,
\big[v_{\,\overline{\nu}_e}\,
\big]_{c\,|\mathbf{p}|\, = \,10\;\text{MeV}}^\text{(exp)}
\,- c\,\Big|\,\Big/c
\lesssim 2 \times 10^{-9}\,.
\eeqa

The challenge is to find an explanation that can help
resolve the apparent discrepancy between these two experimental numbers,
assumed to be correct.
There are, however, two obstacles.
First, there appear to be catastrophic
vacuum-Cherenkov-type energy losses
($\nu_\mu\to \nu_\mu+Z^{0} \to \nu_\mu+e^{-}+e^{+}$)
for the neutrinos
traveling from CERN to the Gran Sasso Laboratory~\cite{CohenGlashow2011}.
Second, similar unacceptable threshold effects may occur for the
pion-decay process ($\pi^{+}\to \mu^{+} +\nu_\mu$)
responsible for the production of the
muon-neutrinos at CERN~\cite{GonzalezMestres2011,Bi-etal2011}.

Both obstacles can be circumvented~\cite{AFKS2011}
if we assume that \emph{deformed} Lorentz
transformations apply to the modified neutrino
dispersion relations, that is, if \emph{relativity still holds}
(further references on deformed Lorentz transformations
and the role of relativity can be found in, e.g.,
Refs.~\cite{AFKS2011,MagueijoSmolin2003}).
In that case, the nonstandard vacuum-Cherenkov-type decay processes
are simply absent, which has been established~\cite{AFKS2011} by
a general argument and a specific calculation in a toy model.
The modification of the pion-decay process
has been argued to be absent as well.

The toy model of Ref.~\cite{AFKS2011}, however,
does not indicate how different neutrino species are to be
treated, let alone the other particles of the standard model.
Here, we implement a suitable modification inspired by
another type of
model~\cite{KlinkhamerPRD2006b,Klinkhamer2011-FPS,KlinkhamerVolovik2011},
in order to find a ``relativistic'' model which satisfies both
\eqref{eq:OPERA-result} and \eqref{eq:SN-bound}.
Henceforth, we set $c=1$.

\section{Neutrino Ansatz}
\label{sec:Neutrino-Ansatz}

The crux of our suggestion is to arrange for
deformed Lorentz transformations which are not mass dependent
as in Ref.~\cite{AFKS2011} but flavor dependent.
(This is perhaps somewhat counterintuitive, but, in the end,
we do not know how the Lorentz deformation arises dynamically.)
Then, the only mixing angles and phase
which enter the game are the standard
ones related to the mass sector
($\theta_{32}$, $\theta_{21}$, $\theta_{13}$, and $\delta$).

Specifically, take the following three neutrino dispersion relations
for the weak-interaction states (flavor label $f=e,\,\mu,\,\tau$):
\beqa\label{eq:disp-rel-flavor-basis}
E^2-p^2- 2\,E^2\,p^2/(\overline{M}_f)^2 &=& (\widetilde{m}_f)^2\,,
\eeqa
with 3-momentum norm  $p\equiv |\mathbf{p}|$
and effective mass values $\widetilde{m}_f$
(in terms of mass eigenvalues $m_n$ and mixing angles
$\theta_{32}$, $\theta_{21}$, and $\theta_{13}$).
As mentioned in the Introduction, the important point
is that relativity still holds, even though
the relevant Lorentz transformations are deformed and flavor dependent.

The particular flavor-dependent generators $\mathcal{N}_j^{\,(f)} $
of Lorentz boosts
can be simply extracted from Eqs.~(8) and (9) in Ref.~\cite{AFKS2011}:
\bsubeqs\label{eq:deformed-Lorentz-boosts}
\beqa
\delta_j^{\,(f)} E&\equiv&\big[\,\mathcal{N}_j^{\,(f)} ,\,E\,\big]
                =p_j\, \Big(1+p^2/(\overline{M}_f)^2 + 2\,E^2/(\overline{M}_f)^2\Big)\,,
\\[2mm]
\delta_j^{\,(f)} p_k&\equiv&\big[\,\mathcal{N}_j^{\,(f)} ,\,p_k\,\big]
                  =\delta_{j,\,k}\, \Big(1-p^2/(\overline{M}_f)^2\Big)\,E\,.
\eeqa
\esubeqs
The left-hand side of \eqref{eq:disp-rel-flavor-basis}
is then invariant to order $1/(\overline{M}_f)^2\,$.
The right-hand side is a scalar
and trivially invariant under these transformations.
By adding higher-order terms to \eqref{eq:deformed-Lorentz-boosts}
the invariance of \eqref{eq:disp-rel-flavor-basis}
can be extended to order $1/(\overline{M}_f)^4$ or
more. For given modified dispersion relation,
the exact Lorentz-boost generators can be obtained by using
the general method of
Ref.~\cite{MagueijoSmolin2003}.
Needless to say, \eqref{eq:disp-rel-flavor-basis} is manifestly
invariant under rotations of the 3-momentum $\mathbf{p}$.
An important consequence of the nonlinear realization
\eqref{eq:deformed-Lorentz-boosts} is that
the addition formula of particle energies is
modified~\cite{AFKS2011,MagueijoSmolin2003}.

 From now on, we neglect neutrino-mass
effects (they will be considered at the end of the paper,
in the Appendix). The model \eqref{eq:disp-rel-flavor-basis},
with all masses $\widetilde{m}_f$ set to zero,
satisfies \eqref{eq:OPERA-result} and \eqref{eq:SN-bound} by taking
\beq\label{eq:Mf-inverse-hierarchies}
(15\,\overline{M}_1)^{-2} \lesssim (\overline{M}_2)^{-2}
\sim (6\;\text{TeV})^{-2}\,,
\eeq
with undetermined $(\overline{M}_3)^{-2}$, for the moment.
Assuming that an interacting theory can be constructed
(a first step will be set in the next section),
the ``Alice-and-Bob'' argument of Sec. II in
Ref.~\cite{AFKS2011} also makes clear that there can be no
vacuum-Cherenkov-type energy losses
in this ``relativistic'' model. This is in contrast to
the vacuum-Cherenkov-type energy losses~\cite{CohenGlashow2011}
in models with a preferred frame (e.g., the dynamical models of
Refs.~\cite{Klinkhamer2011-FPS,KlinkhamerVolovik2011,NojiriOdintsov2011}).

\section{Generalized Ansatz}
\label{sec:v-Ansatz}

As it stands,
the modified dispersion relations \eqref{eq:disp-rel-flavor-basis}
for $\widetilde{m}_{f}=0$ apply only to the neutral
leptons ($\nu_{e},\,\nu_{\mu},\,\nu_{\tau}$), the charged leptons
($e^\pm,\,\mu^\pm,\,\tau^\pm$) having, most likely, more or less
standard Lorentz-invariant dispersion relations.
This implies that the previous discussion holds after the
spontaneous breaking of the $SU(2)_L\times U(1)_Y$ gauge symmetry
(a similar remark applies to a Fermi-point-splitting
model of Lorentz violation~\cite{Klinkhamer2011-FPS}).
We need to find a \mbox{proper gauge-invariant
generalization of \eqref{eq:disp-rel-flavor-basis}.}
Such a gauge-invariant generalization may be one
ingredient of the final interacting theory, which, for the moment,
remains elusive. Anyway, let us start by looking for this one
possible ingredient,
the gauge-invariant generalization of \eqref{eq:disp-rel-flavor-basis}.

Consider the $3\times 15$ Weyl fermions of the standard model (SM)
and three additional right-handed neutrinos
[singlets under $SU(2)$ and $SU(3)$ and with zero $U(1)$ hypercharges],
where all masses are neglected.
These fermions have a species label $a \in \{1, \ldots, 16\}$ and
family label $f \in \{1, 2, 3\}$.
In order to be specific
(and consistent with what was used in the previous section),
we identify $f=1$ with the electron family, $f=2$ with
the muon family, and $f=3$ with the tau family.
The combined label $(a,f)$ will be called the flavor label.

For completeness, we give the SM representations of the
$16$ left- and right-handed Weyl fermions of each family:
\begin{eqnarray}
L &:& \Bigl[\; (3,2)_{1/3} \;\Bigr]_{\rm quarks} +
            \Bigl[\; (1,2)_{-1}  \;\Bigr]_{\rm leptons}
\;,\nonumber\\[2mm]
R &:& \Bigl[\; (3,1)_{4/3}+ (3,1)_{-2/3} \;\Bigr]_{\rm quarks} +
            \Bigl[\; (1,1)_{-2}+ (1,1)_{0}     \;\Bigr]_{\rm leptons}       \;,
\label{SMirreps}
\end{eqnarray}
where the entries in parentheses denote $SU(3)$ and $SU(2)$ irreducible
representations and the suffix the value of the $U(1)$ hypercharge $Y$.
The electric charge $Q$ is given by the combination
$Y/2 + I_3$, with $I_3$ the weak
isospin from the diagonal Hermitian generator $T_3$ of the $SU(2)$
Lie algebra.

For these $48$ Weyl fermions
(labeled by $a \in \{1, \ldots, 16\}$ and $f \in \{1, 2, 3\}$),
the massless dispersion relations  \eqref{eq:disp-rel-flavor-basis}
can be generalized in a relatively simple way:
\bsubeqs\label{eq:disp-rel-flavor-basis-all}
\beqa\label{eq:disp-rel-flavor-basis-all-LV}
E_{a,f}^2-p_{a,f}^2 - 2\,E_{a,f}^2\,p_{a,f}^2/(M_f)^2 &=& 0\,,
\quad\text{for}\;\;
Y_{a,f}\in \{-1,\, -2,\, 0\}\,,
\\[2mm]\label{eq:disp-rel-flavor-basis-all-LI}
E_{a,f}^2-p_{a,f}^2 &=& 0\,,
\quad\text{for}\;\;
Y_{a,f} \notin \{-1,\, -2,\, 0\}\,,
\eeqa
\esubeqs
so that only the leptons have modified
dispersion relations (the quarks have fractional hypercharge values).
The particular generalization presented
in \eqref{eq:disp-rel-flavor-basis-all} is, of course, not unique
and two alternatives will be given later in this section.
Furthermore, there may very well appear
higher-order terms in \eqref{eq:disp-rel-flavor-basis-all-LV}
carrying factors $1/(M_f)^{2 n}$ for $n\geq 2$.

Just as in Sec.~\ref{sec:Neutrino-Ansatz},
the dispersion relations \eqref{eq:disp-rel-flavor-basis-all}
are invariant to order $1/(M_f)^2$ under Lorentz boosts
with the following generators $\mathcal{N}_j^{\,(a,f)}$
[having raised the
flavor label $(a,f)$ for convenience and omitting this label
on the energy $E$ and momentum $p$ in the expressions below]:
\bsubeqs\label{eq:deformed-Lorentz-boosts-all}
\beqa\label{eq:deformed-Lorentz-boosts-all-LV}
\delta_j^{\,(a,f)} E&\equiv&\big[\,\mathcal{N}_j^{\,(a,f)} ,\,E\,\big]
                =p_j\, \big(1+\Delta^{\,(a,f)}\;p^2/(M_f)^2 +\Delta^{\,(a,f)}\; 2\,E^2/(M_f)^2\big)\,,
\\[2mm]
\label{eq:deformed-Lorentz-boosts-all-LI}
\delta_j^{\,(a,f)} p_k&\equiv&\big[\,\mathcal{N}_j^{\,(a,f)} ,\,p_k\,\big]
                  =\delta_{j,\,k}\, \big(1-\Delta^{\,(a,f)}\;p^2/(M_f)^2\big)\,
                   E\,,
\eeqa
using the flavor factor
\beqa
\Delta^{\,(a,f)} &\equiv&
\delta_{\,Y^{\,(a,f)},\,-1}+\delta_{\,Y^{\,(a,f)},\,-2}
+\delta_{\,Y^{\,(a,f)},\,0}\in \{0,\,1\}\,,
\eeqa
\esubeqs
in terms of  a Kronecker-type function for rational numbers
$r,s\in \mathbb{Q}\,$: $\delta_{\,r,s}=1$ for $r=s$ and
$\delta_{\,r,s}=0$ for $r\ne s$.
Hence, standard Lorentz generators apply to fermions with fractional
hypercharge values (quarks) and nonstandard generators to fermions
with integer hypercharge values (leptons).

As the Lorentz invariance of the electron sector has been
verified to a high degree of accuracy, we take
\beq\label{eq:Mf-inverse-hierarchies-all}
0 = (M_1)^{-2} \ll (M_2)^{-2} \sim (6\;\text{TeV})^{-2}\,,
\eeq
where the last approximate equality results from the
experimental input \eqref{eq:OPERA-result}
[the supernova bound \eqref{eq:SN-bound} is trivially satisfied
for $(M_1)^2 \gg (M_2)^2\,$]. The $(M_3)^2$ value is, for the moment,
undetermined.\footnote{It is also conceivable that the leptons in
the tau sector ($f=3$) have \emph{sub}-luminal velocities.
This can be achieved by replacing all occurrences of
$1/(M_3)^2$ in \eqref{eq:disp-rel-flavor-basis-all}
and \eqref{eq:deformed-Lorentz-boosts-all} by $-1/(M_3)^2$,
which corresponds to the first alternative model mentioned
in the discussion below \eqref{eq:disp-rel-flavor-basis-all}.}
For definiteness, we take $(M_3)^2\gtrsim (M_2)^2$.
Furthermore, it is certainly possible to
have $(M_2)^2/(M_1)^2$ not zero as in
\eqref{eq:Mf-inverse-hierarchies-all} but sufficiently small.
Still, setting $(M_2)^2/(M_1)^2=0$ focusses the discussion
on the muon sector.

The model \eqref{eq:disp-rel-flavor-basis-all}--\eqref{eq:Mf-inverse-hierarchies-all}
can only correspond to an effective theory valid for energies
below $\text{min}(M_1,\,M_2,\,M_3)=
\text{O}(10\;\text{TeV})$. It remains to be seen what the
theory looks like at energies $E \gtrsim M_2 \sim 10\;\text{TeV}$.
Note that if \eqref{eq:OPERA-result} were to be
reduced by a
factor $10^{16}$,  
the required value of $M_2$ would be increased by a
factor $10^{8}$,  
giving a scale $M_2\sim 10^{12}\;\text{GeV}$.
Considering such large scales, there is a second alternative model:
all 48 Weyl fermions have the same modified
dispersion relation  \eqref{eq:disp-rel-flavor-basis-all-LV}
with $M_1=M_2=M_3\equiv M \gg 10\;\text{TeV}$
(or perhaps $M \sim 10\;\text{TeV}$ after all?).
The interacting theory may be easier to construct
for this alternative model with all fermions treated equally.
An entirely different model which circumvents the interaction
problem is discussed in the Appendix.

Returning to model
\eqref{eq:disp-rel-flavor-basis-all}--\eqref{eq:Mf-inverse-hierarchies-all},
the OPERA result \eqref{eq:OPERA-result} then predicts a matching
relative change of the muon velocity:
\beqa\label{eq:v-muon-prediction}
\Big(
\big[v_{\,\mu^\pm}\,
\big]_{c\,|\mathbf{p}|\, = \,10\;\text{GeV}}^\text{(model)}
-c\Big)\Big/c \sim 10^{-5}.
\eeqa
Again, indirect bounds on $|v_{\mu^\pm}/c-1|$
relying on modified decay processes~\cite{Altschul2007}
may not apply to this ``relativistic'' model.
But there are also direct laboratory experiments,
for example, experiments to measure
the muon anomalous magnetic moment~\cite{Bennett-etal2007,PDG2010}.
It seems unlikely that per-mill changes in the muon velocity
at energies of order $10^2\;\text{GeV}$ would have gone unnoticed,
but this needs to be checked (in a theoretical framework without
preferred frame and with appropriate interactions).

\section*{ACKNOWLEDGMENTS}
\noindent It is a pleasure to thank L. Smolin for helpful discussions.

\begin{appendix}
\section{STERILE-NEUTRINO ANSATZ}
\label{sec:App}

It may turn out to be impossible to embed the Lorentz violation of
\eqref{eq:disp-rel-flavor-basis-all} into an interacting theory
containing the SM particles
and, simultaneously, avoid excessive energy
losses~\cite{CohenGlashow2011} from the vacuum-Cherenkov-type process
$\nu_\mu\to \nu_\mu+Z^{0} \to \nu_\mu+e^{-}+e^{+}$.
In that case, there exists an alternative approach, which, however,
looses the prediction \eqref{eq:v-muon-prediction}.
This approach allows us to include neutrino-mass effects,
which were neglected in the discussion of the main text.
In fact, the neutrino eigenstates considered in the present appendix
(labeled by $n\in\mathbb{N}$) are those of mass.

The point is that the modified dispersion
relation \eqref{eq:disp-rel-flavor-basis} can also apply to
a light sterile neutrino, which has been invoked to
explain~\cite{Giudice-etal2011,HannestadSloth2011,Nicolaidis2011}
the OPERA result \eqref{eq:OPERA-result}.
The three standard neutrinos ($n=1,\,2,\,3$) then travel
with luminal or subluminal velocities,
while the light sterile neutrino (mass $m_4 \sim 0.1\,$--$\,1\;\text{eV}$)
is superluminal.
The velocity `$v$' appearing in \eqref{eq:OPERA-result} is
an average of these sub- and superluminal velocities with
weights depending on the mixing angles (see below).
The superluminal sterile neutrino, by definition, does not couple
to the $Z^{0}$ boson and does not suffer from catastrophic
energy losses~\cite{CohenGlashow2011}.
For further details on and challenges for such a type of
phenomenological superluminal-sterile-neutrino model,
see Refs.~\cite{Giudice-etal2011,HannestadSloth2011,Nicolaidis2011,Winter2011}
and references therein.

The fundamental question, of course, is how the light sterile neutrino
acquires a superluminal maximum velocity.
One possible explanation relies on the introduction
of one or more extra spatial dimensions
and a braneworld with appropriate
metric~\cite{HannestadSloth2011,Nicolaidis2011,Pas-etal2005}.

We suggest, instead, that the theory remains four-dimensional
but has two sectors without direct interactions, one containing
all standard-model particles (fermions, as well as gauge and Higgs bosons)
with the standard linear Lorentz transformations and another
containing gauge-singlet (sterile) particles with
deformed nonlinear Lorentz transformations
as discussed in Sec.~\ref{sec:Neutrino-Ansatz}.
Taking over the previous example \eqref{eq:disp-rel-flavor-basis}
and temporarily reinstating the light velocity $c$,
the dispersion relations of the four neutrino mass eigenstates
($n=1,\,2,\,3,\,4$) are given by
\bsubeqs\label{eq:disp-rel-mass-basis}
\beqa
E^2-c^{\,2}\,p^2- 2\,E^2\,c^{\,2}\,p^2\big/\big(\overline{M}_n\,c^{\,2}\big)^2
&=& \big(m_n\,c^{\,2}\big)^2\,,\\[2mm]
1/\overline{M}_1=1/\overline{M}_2=1/\overline{M}_3&=& 0\,,
\eeqa
\esubeqs
with $\overline{M}_4=\text{O}(10\;\text{TeV}/c^{\,2})$
if OPERA's result \eqref{eq:OPERA-result} sets the scale
and if there is significant mixing between the $f=\mu$ flavor state
and the $n=4$ mass state.
The dynamical origin of this active-sterile
neutrino mixing certainly needs to be clarified.
Perhaps there exists an indirect (nonperturbative) gravitational
interaction between the two sectors resulting in an effective
contact-interaction term in the action, which is suppressed by
at least one factor $1/E_\text{Planck}$
(cf. Ref.~\cite{Berezinsky-etal2003} and references therein).

Alternatively, genuine (preferred-frame) Lorentz violation
of the light sterile neutrino may come from the Fermi-point-splitting
mechanism~\cite{Klinkhamer2011-FPS}
or from the spontaneous breaking of Lorentz invariance
(SBLI)~\cite{KlinkhamerVolovik2011}, both operating
in the known $3+1$ spacetime dimensions.

The SBLI explanation, in particular, is quite attractive as it only
relies on the appearance of a fermion condensate,
possibly coming from the multifermion
interaction (7) of Ref.~\cite{KlinkhamerVolovik2011}
applied to the gauge-singlet field of the sterile neutrino. With a
timelike fermion-condensation vector $(b_\alpha)=(b_0,\,0,\,0,\,0)$
as discussed in Ref.~\cite{KlinkhamerVolovik2011} and
$c$ temporarily reinstated,
the dispersion relations of the four neutrinos are given by
\bsubeqs\label{eq:disp-sbli-mass-basis}
\beqa
E^2 &=& c^{\,2}\,p^2+\big(m_n\,c^{\,2}\big)^2\,,
\quad\text{for}\;n=1,\,2,\,3,
\label{eq:disp-sbli-mass-basis-normal}\\[2mm]
E^2 &\sim& c_4^{\,2}\,p^2+\big(m_4\,c_4^{\,2}\big)^2\,,
\quad\text{for}\;n=4,
\label{eq:disp-sbli-mass-basis-sterile}\\[2mm]
c_4 &\equiv& \big[1-(b_0)^2\,\big]^{-1/2}\;c\,,
\label{eq:disp-sbli-mass-basis-def-c4}\\[2mm]
m_4 &\equiv& \big[1-(b_0)^2\,\big]^{1/2}\;\widehat{m}_4\,,   
\label{eq:disp-sbli-mass-basis-def-m4}
\eeqa
\esubeqs
where $\widehat{m}_4$ is the sterile-neutrino mass without
spontaneous symmetry breaking (order parameter $b_0=0$)
and where \eqref{eq:disp-sbli-mass-basis-sterile} holds
for $E\gg |\widehat{m}_4\,c^{\,2}/b_0|$, corresponding to $E\gg 10^{2}\;\text{eV}$
for $|b_0|\sim 10^{-2}$ and $m_4 \sim \text{eV}/c^{\,2}$.
The order of magnitude of $|b_0|$ is indeed $10^{-2}$ if OPERA's
result \eqref{eq:OPERA-result} sets the scale
and if there is significant $\mu 4$ mixing
(a somewhat larger value of $|b_0|$ can compensate for a somewhat smaller
value of the mixing angle $\theta_{\mu 4}\,$; see also the discussion in
Refs.~\cite{HannestadSloth2011,Winter2011}).\footnote{For
the CERN--GranSasso
(CNGS) setup, a narrow symmetric pulse of nearly mono-energetic
muon-neutrinos produced at CERN
(cf. Sec.~9 of Ref.~\cite[(b)]{OPERA2011})
would then give a broadened, possibly asymmetric pulse
of muon-neutrinos detected by OPERA in the Gran Sasso Laboratory.
Detailed measurements of the final
pulse profile could rule out (or confirm)
the sterile-neutrino hypothesis.}

The modified dispersion relation
shown in \eqref{eq:disp-sbli-mass-basis-sterile}
is only the simplest one possible.
By considering higher-derivative terms in the action,
SBLI can also give nonstandard terms $p^{2n}$ with $n\geq 2$
on the right-hand side of \eqref{eq:disp-sbli-mass-basis-sterile}.
The resulting energy dependence of the velocity `$v$'
in the theoretical counterparts of
Eqs.~\eqref{eq:OPERA-result} and \eqref{eq:SN-bound},
together with equal mixing of the sterile neutrino
and the three flavors of active neutrinos
(e.g., mixing angles $\theta_{e 4}=\theta_{\mu 4}=\theta_{\tau 4}$),
make for phenomenologically attractive models,
provided the leakage of Lorentz violation
to the charged-lepton sector by quantum effects
(e.g., loop corrections to the electron propagator)
can be kept small enough~\cite{Giudice-etal2011}.

The extra-dimensional braneworld explanation
for the superluminal velocity of the sterile neutrino
suggests an equal superluminal velocity of gravitational waves,
which may already be in conflict with existing
bounds~\cite{HannestadSloth2011}.
Now, compare this expected behavior with that of the
three four-dimensional mechanisms mentioned above.
For the two-sector explanation, it is not really
clear if the gravitational-wave velocity is modified or not.
But, for the Fermi-point-splitting
and SBLI explanations in their basic form,
the gravitational-wave velocity is definitely equal
to the speed of light, $c$.

As a possible explanation of the OPERA result~\cite{OPERA2011},
the simplest version of a
superluminal-sterile-neutrino model~\cite{Giudice-etal2011,HannestadSloth2011}
is perhaps one with SBLI~\cite{KlinkhamerVolovik2011}.
Further sterile-neutrino-SBLI models are discussed
in Ref.~\cite{Klinkhamer2011-sterile-4D}.
\end{appendix}


\end{document}